# Ferroelectric control of the spin texture in germanium telluride

C. Rinaldi[1,2*], S. Varotto[1], M. Asa[1], J. Slawinska[3], J. Fujii[4], G. Vinai[4], S. Cecchi[5], R. Calarco[5],

I. Vobornik[4], G. Panaccione[4], S. Picozzi[3], R. Bertacco[1,2*]

[1]Department of Physics, Politecnico di Milano, 20133 Milano, Italy.
[2]IFN-CNR, c/o Politecnico di Milano, 20133 Milano, Italy.
[3]Consiglio Nazionale delle Ricerche CNR-SPIN, Sede Temporanea di Chieti, c/o Univ. "G. D'Annunzio", 66100 Chieti, Italy
[4]Istituto Officina dei Materiali CNR Laboratorio TASC, AREA Science Park - Basovizza, 34149 Trieste, Italy.
[5]Paul-Drude-Institut für Festkörperelektronik, Hausvogteiplatz 5-7, 10117 Berlin, Germany

*Correspondence to: riccardo.bertacco@polimi.it; christian.rinaldi@polimi.it

## Abstract

The electrical manipulation of spins in semiconductors, without magnetic fields or auxiliary ferromagnetic materials, represents the "holy grail" for spintronics. The use of Rashba effect is very attractive because the *k*-dependent spin-splitting is originated by an electric field. So far only tiny effects in two-dimensional electron gases (2DEG) have been exploited. Recently, GeTe has been predicted to have bulk bands with giant Rashba-like splitting, originated by the inversion symmetry breaking due to ferroelectric polarization. In this work we show that GeTe(111) surfaces with inwards or outwards ferroelectric polarizations display opposite sense of circulation of spin in bulk Rashba bands, as seen by spin and angular resolved photoemission experiments. Our results represent the first experimental demonstration of ferroelectric control of the spin texture in a semiconductor, a fundamental milestone towards the exploitation of the non-volatile electrically switchable spin texture of GeTe in spintronic devices.

**Introduction**

While Moore's law seems to approach its limit of validity, the search for new paradigms allowing to further improve the computing capabilities of electronic devices is entering a final rush. Spintronics is a promising route in this perspective, but so far its success stories are limited to the field of memories. To enter the area of computing, devices capable of manipulating the information encoded in the spin are needed. In this sense, a lot of effort is currently carried out in the fields of spin logic,[1] magnon spintronics,[2,3] and semiconductor spintronics.[4–7] Nevertheless, about twenty years after the pioneering idea of "spin transistor" proposed by Datta and Das,[8] many practical limitations still prevent its implementation.[9] The dream remains to manipulate spins within semiconductors devices, to exploit the full potential of materials with a gap for charge control, without use of auxiliary ferromagnetic materials and magnetic fields. Beyond magnetic semiconductors,[10] which still suffer from low temperature operation and the need of external magnetic fields to control the spin texture, new materials and concepts are necessary. In this context, the recently introduced class of FerroElectric Rashba SemiConductors (FERSCs),[11] whose father compound is α-GeTe,[12] is highly promising. They are semiconductors and also ferroelectrics, so that the remanent ferroelectric polarization vector breaks the inversion symmetry and determines a giant bulk Rashba $k$-dependent spin-splitting of the bands.[13] Remarkably, density functional theory (DFT) simulations predict that the spin direction in each sub-band should reverse upon inversion of the ferroelectric polarization, thereby allowing its electrical control. In perspective, these unique features could be exploited in novel devices[14,15] integrating memory and computing functionalities within the very same channel of a spin-transistor.[16]

In this paper we address a fundamental issue on the roadmap towards the exploitation of FERSCs: the electric switchability of the spin texture in α-GeTe. The switching of the ferroelectric (FE) polarization has been already demonstrated in GeTe(111) films[13] and more recently in GeTe nanowires.[17] On the other hand, an experimental proof of the reversal of the spin texture (clockwise or counterclockwise sense of circulation of spins in bulk Rashba-like bands) for opposite FE polarization is still missing. So far, only a link between the direction of the FE polarization and the spin orientation in the surface Rashba sub-bands has been reported.[13] More recently, the impact of magnetic fields on the spin texture of Mn-doped GeTe films has been investigated, but always for fixed FE polarization.[18] Here we provide evidence for the ferroelectric control of spin texture, by using spin and angular resolved photoemission spectroscopy (S-ARPES) on GeTe(111) surfaces with opposite FE polarization. Our findings demonstrate the feasibility of the electric switching of spin texture in ferroelectric Rashba semiconductors, thus opening the route to their exploitation in spintronic devices with pure electric control of their operation.

**Engineering of GeTe(111) surfaces**

To avoid the use of a top electrode for in-situ application of voltage pulses, which would create serious problems for highly surface sensitive and time demanding experiments such as S-ARPES, we developed a method for preparing α-GeTe(111) surfaces with opposite FE polarization, associated to different terminations. Rhombohedral α-GeTe(111) results from the stacking of Ge and Te planes which are not equidistant and thus give rise to a net electric dipole, being Te more electronegative than Ge.[19,20] As the energetically favored termination is generated by the breaking of long (rather than short) bonds, a Te terminated surface is expected to display a dipole pointing outwards ($P_{out}$) while a Ge terminated one will have a net dipole inwards ($P_{in}$), as depicted in panels (a) and (a') of Figure 1. Even though calculations predict the Te-terminated GeTe(111) to be more stable than the Ge-terminated one by 60 meV Å$^{-2}$, this surface energy difference is largely reduced by the presence of reconstructions,[21] vacancies and other kind of defects, thus suggesting the possibility of stabilizing both terminations in real surfaces.

GeTe(111) films, 23 nm thick, were grown by Molecular Beam Epitaxy (MBE) on Si(111) and then capped with 20 nm of Te to prevent contamination due to exposure to atmosphere. A reliable protocol for controlled Te-desorption in ultra-high vacuum (UHV) has been first optimized on-campus to obtain Ge- and Te-rich surfaces with opposite polarization, as checked by XPS with Al-Kα radiation and Piezoresponse Force Microscopy (PFM) afterwards. After insertion in vacuum at the APE beamline, the samples were annealed at about 240 °C and 260 °C for one hour, controlling the heater current. According to the recipe previously optimized, we obtained a first sample ($S_{Te}$) with a Te-rich surface and a second one ($S_{Ge}$) with a Ge-rich surface, as confirmed by the in-situ analysis of the XPS spectra taken at 800 eV photon energy. In fact, from the relative intensity of the Ge 3d and Te 4d peaks, normalized to the analyser transmission and tabulated atomic photoemission cross sections, the average stoichiometries of $S_{Te}$ and $S_{Ge}$ result to be $Ge_{0.39}Te_{0.61}$ and $Ge_{0.46}Te_{0.54}$ respectively. The uncertainty on the relative stoichiometry is ±0.02 and it mainly arises from the error on the estimation of the peaks' area. Within the photoemission probing depth (~17 Å at 800 eV photon energy) $S_{Te}$ is clearly Te-rich, while the Ge concentration in $S_{Ge}$ is slightly above that of GeTe films, which typically displays 10% Ge vacancies ($Ge_{0.45}Te_{0.55}$),[22,23] thus pointing towards a Ge enrichment of $S_{Ge}$. Data taken with Al-Kα radiation (1486.6 eV), after the beamtime, confirm that $S_{Te}$ ($S_{Ge}$) is Te (Ge) rich at surface. Furthermore, the deconvolution of XPS spectra, using surface and bulk components, shows that the Te (Ge) enrichment is localized at surface. Based on these considerations, in the following we will compare our ARPES data from $S_{Te}$ and $S_{Ge}$ with Density Functional Theory simulations of Ge- and Te-terminated surfaces. In fact, these truncated bulk systems represent the simplest models of Te- and Ge-rich surfaces with outwards and inwards FE polarization.

**Ferroelectric state of engineered surfaces**

After the beamtime we investigated the virgin ferroelectric state of the two samples by PFM. In Figure 1b (1b') we report the phase signal recorded on $S_{Te}$ ($S_{Ge}$) after writing a large square with + 7 V (-7 V) bias on the AFM tip and then an inner square with opposite bias. FE patterns are stable over more than 24 hours, thus indicating the robustness of ferroelectricity in GeTe films. The inner square displays the same contrast (PFM phase) of the unpoled area, indicating an outwards (inwards) virgin ferroelectric polarization in $S_{Te}$ ($S_{Ge}$). To confirm these findings, we measured the virgin curve and the full ferroelectric hysteresis loops by sweeping the PFM tip voltage. Two characteristic loops measured on $S_{Te}$ and $S_{Ge}$ are reported in panels (c) and (c') of Figure 1, as representative of the average response of the entire sample area probed by S-ARPES. The virgin curve measured for $S_{Te}$ indicates the initial state is close to that obtained for negative saturating voltages, while the opposite holds for the sample $S_{Ge}$. This is a clear indication that $S_{Te}$ and $S_{Ge}$ present two opposite outwards and inwards virgin FE polarizations. These samples are ideal candidates for investigating the connection between FE polarization and spin texture.

**Surface and bulk-like bands in Te (Ge) rich surfaces**

In Figure 2 we show the bands dispersion and the corresponding spin texture calculated by DFT for Te- and Ge-terminated GeTe(111) surfaces displaying respectively $P_{out}$ and $P_{in}$ FE polarization, that will be compared with S-ARPES from $S_{Te}$ (Te-rich surface) and $S_{Ge}$ (Ge-rich surface), respectively. The slabs used for simulations are the same reported in Figure 1 without Ge vacancies. The bulk high symmetry directions ZU and ZA, together with the corresponding surface directions $\overline{\Gamma M}$ and $\overline{\Gamma K}$, are shown in panel 2d. The FE polarization is parallel to ΓZ direction of the reciprocal space, i.e. the (111) direction of the crystal. Band dispersions along high symmetry directions are presented in panel 2a for $S_{Te}$ and 2a' for $S_{Ge}$, after projection of the spectral function on the surface layers and on the bulk in the semi-infinite model, to single out surface and bulk-like contributions. The spin texture is reported in panel b for $S_{Te}$ and b' for $S_{Ge}$, where the non-null spin components perpendicular to the wave vector are shown along the high symmetry bulk (surface) ZA ($\overline{\Gamma K}$) and ZU ($\overline{\Gamma M}$) directions. While the shape of bulk Rashba sub-bands is not affected by FE polarization reversal, their spin texture is reversed, according to the main concept of FERSCs. This is evident from the comparison of the isoenergy cuts, taken at 0.5 eV below the top of the valence band, reported in panels c and c', where arrows indicate the local spin direction. On the other hand, surface bands with Rashba splitting are very different for the two termination. In the Te-terminated one ($P_{out}$) they display a clear Rashba-like splitting and cross the Fermi energy at higher wave vectors with respect to the bulk bands. In the Ge-terminated one ($P_{in}$), instead, the Rashba splitting of the surface bands in the gap is largely suppressed and surface bands shift towards the conduction band, without crossing the Fermi level at high momenta.

The remarkable difference between surface Rashba like bands predicted by DFT for Te- and Ge-termination has an experimental counterpart in ARPES data reported in Figure 3 for $S_{Te}$ (panels a-h) and $S_{Ge}$ (a'-h'). Panels 3a and 3b present the experimental band dispersions along ZA and ZU for $S_{Te}$. Corresponding isoenergy cuts at 0, 0.25 and 0.5 eV BE in panels 3c, 3d, 3e are compared with theoretical ones for a Te-terminated surface. In the following, we will use a simplified distinction between "surface" and "bulk" Rashba states. Having in mind that ARPES at 20 eV probes just a few atomic layers underneath the sample surface, we identify as bulk states those displaying a sizable photon energy or $k_z$ dispersion.[13] Rigorously, these are not true bulk states, but can be viewed as surface-bulk resonances[24] or simply states with sizable projection on bulk states, so as they mainly reflect the bulk behaviour.[25]

Both band dispersions and isoenergy cuts from sample $S_{Te}$ are very similar to data available in the literature for Te-terminated α-GeTe(111).[13,24,25] In agreement with DFT calculations, prominent surface bands with Rashba splitting ($S_1$, $S_2$) are seen at large momenta, especially along the ZA direction. Two outer spin split bands with six-fold symmetry and "arms" along the equivalent ZU directions are clearly visible in the isoenergy cuts of panels 3c, 3d (marked by green ticks). However, already at 0.25 eV BE (Figure 3d), an inner six-fold star (orange ticks) appears, rotated by 30 degrees with respect to the surface one, i.e. with arms along ZA. This is ascribed to the bulk-like bands ($B_1$ and $B_2$) evident in the band dispersion along ZU of panel 3b. The isoenergy cut at 0.5 eV (Figure 3e), instead, mainly reflects the symmetry of the bulk inner star, because at this BE the cut of states with surface character occurs at higher momenta.

The scenario for the $S_{Ge}$ sample is completely different because surface states are almost absent, in agreement with DFT simulations. Indeed the prominent surface Rashba bands $S_1$, $S_2$ along ZA in Figure 3a are missing in 3a', while bulk bands $B_1$, $B_2$ along ZU[12] are similar in panels 3b and 3b'. The absence of $S_1$ and $S_2$ surface states is even more evident from the Fermi energy cut of panel 3c', which does not display the outer six-fold double star of panel 3c. Besides, the isoenergy cut at 0.25 eV (panel 3d') already reflects the symmetry of bulk states, like the inner star in Figure 3d-e from sample $S_{Te}$, i.e. with arms along ZA.

To summarize, ARPES data from samples $S_{Te}$ and $S_{Ge}$ show band dispersions in good agreement with those calculated for a Te-terminated ($P_{out}$) and Ge-terminated ($P_{in}$) surface, respectively. This represents a self-consistent proof of the reliability of our method for preparing GeTe(111) surfaces with opposite FE polarization.

**Spin texture**

Figure 4 reports spin-resolved ARPES data from $S_{Te}$ and $S_{Ge}$. Here we focus on the connection between the spin texture of bulk Rashba bands and the FE polarization, which is the key concept of FERSC materials. In fact, the Rashba splitting of surface Rashba bands can be largely affected, or even suppressed, by proximity with other materials in a multilayer,[26] or by the surface electric field due to screening charges. In sample $S_{Te}$, which displays both surface and bulk Rashba bands, we performed spin polarized scans at fixed momenta ($\mathbf{k}_1$, $-\mathbf{k}_1$) marked in panels 4f and 4g, along the equivalent ZU direction at 30 degrees with respect to $k_x$. Even though these are not the points where the Rashba splitting is maximized, for $\pm\,\mathbf{k}_1$ only bulk bands $B_{1,2}$ are expected to contribute to the photoemission signal at BE greater than 0.2 eV (see Figure 2a and 2b). The spin polarized spectra and corresponding spin-polarization are reported in panels 4b and 4c for $\mathbf{k}_1$, 4d and 4e for $-\mathbf{k}_1$. With reference to the polarimeter quantization axis set along the negative direction of $k_y$, at $\mathbf{k}_1$ we find a positive peak in the spin polarization at about 0.2 eV and a negative one at about 0.5 eV (panel c), corresponding to the crossing of the outer and inner band $B_1$ and $B_2$, respectively. The opposite occurs at $-\mathbf{k}_1$ as expected for GeTe Rashba bands.[12,24,25] The sense of circulation of spins resulting from our data is sketched in Figures 4f and 4g, by arrows superimposed to the isoenergy cuts taken at 0.18 eV and 0.5 eV. In agreement with DFT calculations, for a Te-terminated ($P_{out}$) surface the sense of circulation of spins is clockwise for the outer band and counterclockwise for the inner one.

For $S_{Ge}$, the analysis of the spin texture of bulk bands is simpler, due to the lack of surface bands. In this case, we choose opposite $k$ points ($\mathbf{k}_2$, $-\mathbf{k}_2$), along $k_y$ (ZU direction), where the maximum Rashba energy splitting ($E_R$) of bulk bands $B_{1,2}$ is expected (see Figure 4a' and 2a'). The quantization axis of the spin polarimeter was set orthogonal to the wave vectors, towards the positive $k_x$ direction. Spin polarized spectra in Figure 4b' display two prominent peaks with opposite spin, arising from the crossing of $B_1$ and $B_2$ bands at $\mathbf{k}_2$. Their energy splitting of about 200 meV is in good agreement with the expected value of the Rashba energy $E_R$, according to theoretical predictions[12] and recent experimental findings[24,25]. Noteworthy, the sign of the spin polarization of the two peaks reverts when moving from $\mathbf{k}_2$ to $-\mathbf{k}_2$, as it appears from the comparison of panels 4b', 4c' and 4d', 4e'. To determine the sense of circulation of spins in the outer and inner bands we simply note that in panel (c') the peak at lower BE (outer band) has a negative polarization with respect to the quantization axis, i.e. the spin is directed along the negative direction of $k_x$ (counterclockwise rotation). The opposite holds for the peak at higher BE (inner band), so that the sense of circulation of the spin there is clockwise. The corresponding spin texture is sketched in panels 4f', and 4g'.

Crucial for FERSC demonstration, the sense of circulation of spin in the inner and outer bands is opposite in samples $S_{Te}$ (Figure 4g) and $S_{Ge}$ (Figure 4g'), which display outward and inward FE polarization respectively. This means that the spin texture is locked to the FE polarization since it reverts when the FE polarization is switched.

To summarize, we discussed on the ferroelectric control of the spin texture in GeTe, as originally predicted by Di Sante *et al.*.[12] Using a surface engineering strategy to prepare two high quality surfaces with opposite polarizations, we demonstrated that the spin texture in GeTe is locked to the ferroelectric state. The latter can be easily manipulated, as stable FE domains of both polarities can be electrically written and erased at will, thus providing a reliable way to act on the former. Our findings indicate that a full electric control of the spin in a semiconductor is feasible, without magnetic fields and/or adjacent magnetic layers. This represents a fundamental achievement towards the deployment of GeTe in spintronic devices exploiting the rich physics of Rashba effect and the additional degree of freedom arising from the electric reconfigurability of the spin texture.

**Methods**

**Sample preparation**

GeTe films were epitaxially grown by MBE on Si(111) substrates (p-type B-doped, resistivity 1–10 Ω cm, miscut < 0.1 degrees, 100 nm thermal oxide capping layer).[27] The substrates were wet cleaned before loading them into the MBE system and thermal desorption of water was induced in the growth chamber by annealing. An appropriate treatment allows to obtain a Si(111)-($\sqrt{3}\times\sqrt{3}$))R30°-Sb surface[28] before cooling down to the deposition temperature at 250 °C. The deposition was performed using Ge and Te dual-filament effusion cells. A Te/Ge flux ratio of ~1.6 was used. After deposition of 23 nm of GeTe, a 20 nm thick capping layer of amorphous Te was deposited on top, in order to prevent contamination upon exposure to atmosphere. To prepare a clean and well-ordered GeTe(111) surface, a controlled thermal desorption process in UHV was optimized. An annealing of 1 hour at about 240°C produces a Te-rich surface displaying a pristine polarization $P_{out}$, while the same annealing at 260°C causes a complete desorption of Te (the more volatile species) and produces a Ge-rich surface with $P_{in}$.

**PFM analysis**

Piezoresponse Force Microscopy was performed using a Keysight 5600LS atomic force microscope operating in single-frequency excitation mode. Spectroscopy and imaging were performed using conductive tips by Applied NanoStructures Inc. (AppNano ANSCM-PT, highly n-doped single crystal Si coated with Pt, $L$= 225 µm, $k$= 3 N/m). The typical driving amplitude was 2 $V_{AC}$ and the driving frequencies were between 20 or 250 kHz depending on the specific frequency response. Due to the tiny piezoelectric coefficient $d_{33}$ in GeTe, 1 pm/V[29] to be compared with ~300 pm/V for PZT[30], PFM was performed in Switching Spectroscopy mode (SS-PFM[31] or Pulsed DC Mode[32]), by applying dc voltage pulses and measuring the piezoelectric signal at zero dc bias. The technique avoids the presence of spurious electrostatic effects such as the cantilever-sample capacitive force and enables measurements on materials with a relatively small piezoelectric response.

Finally, the uniformity of samples was checked by acquiring the hysteresis loop on many locations (~10-100 points) over the whole sample surface.

**DFT simulations**

Our DFT calculations were performed by using the GREEN code[33] interfaced with the SIESTA package.[34] The exchange and correlation terms were considered within the GGA in the Perdew–Burke–Ernzerhof formalism.[35] Core electrons were replaced by norm-conserving pseudopotentials of the Troulliers-Martin type. The atomic orbital (AO) basis set consisted of double-zeta polarized numerical orbitals strictly localized. The confinement energy in the basis generation process was set to 100 meV. Real space three-center integrals were computed over 3D-grids with a resolution equivalent to 1000 Rydbergs mesh cut-off. Spin–orbit coupling has been self-consistently taken into account as implemented in Ref.[36]. Surface slab models were constructed as five hexagonal unit cells of α-GeTe, stacked along the [0001] direction with an additional Te (Ge) layer to simulate two unrelaxed Te- (Ge-) terminated surfaces (31 atoms in the unit cell) with different orientation of polarization (outwards and inwards). For all calculations, Brillouin zone (BZ) integrations were performed over $k$-supercells of (20 × 20) while the temperature $k_BT$ in the Fermi–Dirac distribution was set to 10 meV. Dipole-dipole interactions among the neighbouring supercells were suppressed via the usual dipole corrections applied along the $z$-direction. The electronic and spin structures have been calculated in the form (**k**, $E$)-resolved projected density of states PDOS and density of magnetization vector **m** employing the Green's functions matching technique described in details elsewhere[37].

**Angular resolved photoemission spectroscopy**

ARPES and spin polarized ARPES (S-ARPES) spectra have been measured at the low energy branch of APE-NFFA beamline of Electra, using the Omicron-Scienta DA30 electron energy analyser that operates in the deflection mode. The beamline provides the photons in the VUV range with variable polarization; the data presented in the manuscript were acquired with the linearly polarized photons with the polarization vector lying in the scattering plane. The analyser is further equipped with two highly efficient spin polarimeters,[38] for the determination of spin polarization vector. The polarimetry is based on the low energy electron scattering from the magnetic Fe(001)-p(1x1) target,[39,40] whose magnetization can be reversed along the two easy orthogonal directions for the determination of three-dimensional spin polarization. The spin asymmetry of this scattering, equivalent to the Sherman function of Mott polarimeters, is 25%. This value was used to renormalize spectra and obtain the spin polarization of photoelectrons. The energy and angular resolution without (with) spin analysis were 40 meV (100 meV) and <0.2° (<1.5°). All measurements have been performed at liquid nitrogen temperature.

**Acknowledgements**

We are grateful for helpful discussions with G. Rossi and M. Cantoni. C.R., S.V., M.A. and R.B. acknowledge financial support by the Cariplo Foundation, grant n. 2013-0726 (MAGISTER) and grant no. 2013-0623 (SEARCH IV). This work has been partly performed in the framework of the nanoscience foundry and fine analysis (NFFA-MIUR Italy) project. R.C. and S.C. thank S. Behnke and C. Stemmler for technical support at the MBE; the Leibniz Gemeinschaft within the Leibniz Competition on a project entitled: "Epitaxial phase change superlattices designed for investigation of non-thermal switching" for partial funding.

**Figures and legends**

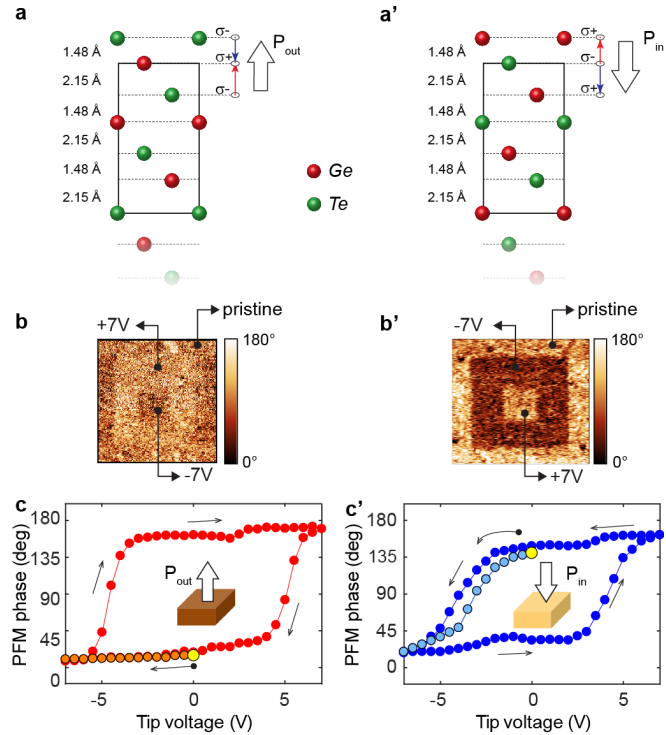

**Figure 1. GeTe(111) surfaces with opposite FE polarization. a, a',** Sketch of the Ge and Te planes for the Te- and Ge-termination, respectively. The distances given on the left refer to the unrelaxed interlayer distances. Only topmost surface atoms of the slab are shown. The black rectangle denotes a bulk hexagonal unit cell used as a building block to construct the (111) surface in DFT calculations. The net FE polarization $P_{out}$ ($P_{in}$) (white arrows) arises from the interatomic dipoles shown with blue and red arrows. **b, b',** Piezoresponse phase images recorded on $S_{Te}$ ($S_{Ge}$) after poling with the tip at +7 V (-7 V) and -7 V (+7 V) over two concentric squares of 1.5 and 0.5 µm side. **c, c',** PFM phase signal showing the pristine polarization state and the ferroelectric hysteresis loop, as measured ex-situ on $S_{Te}$ and $S_{Ge}$ after the S-ARPES experiment. The controlled thermal desorption of the Te capping layer leads to a virgin state FE polarization $P_{out}$ and $P_{in}$ in the two samples, respectively.

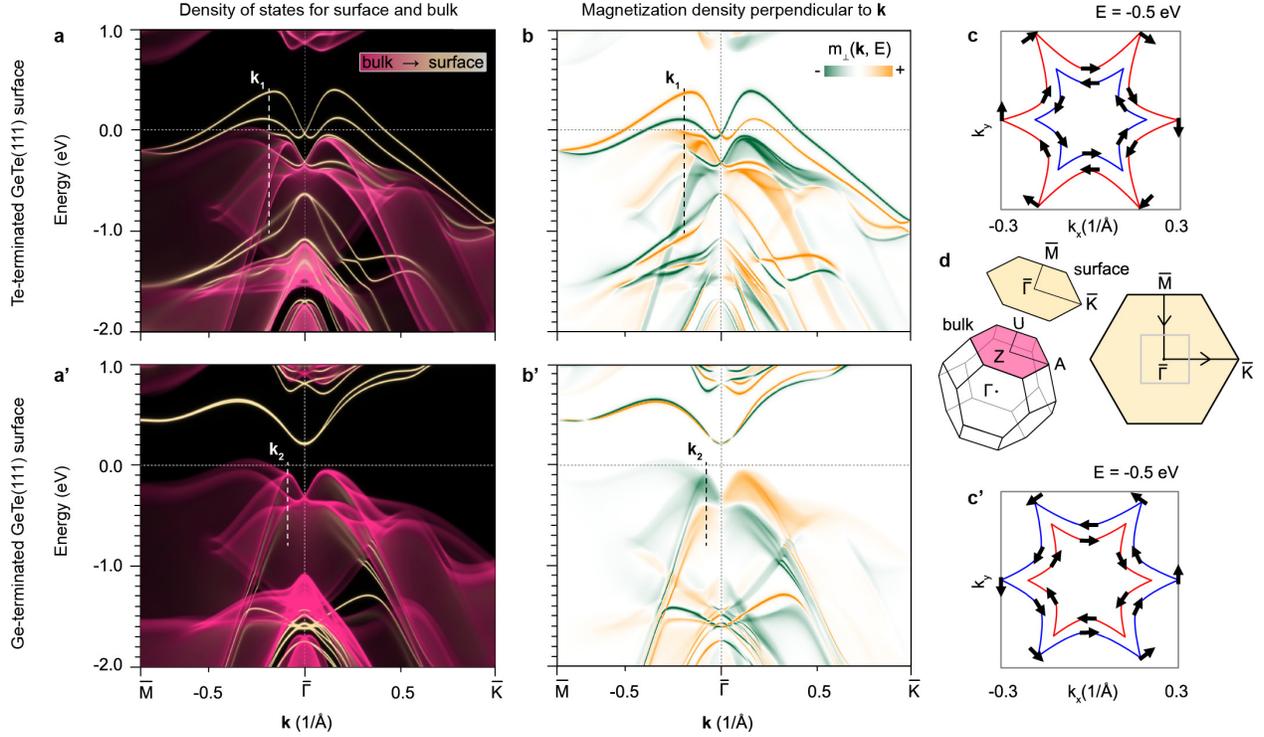

**Figure 2. DFT calculations of GeTe(111) surfaces with different terminations. a,** Density of states (Spectral function) of Te-terminated GeTe(111) surface with outwards polarization, projected on bulk (deep pink) and surface (yellow) principal layers calculated via Green's functions technique for the semi-infinite model of the surface. Brighter tones of pink (yellow) bands indicate higher intensity of bulk (surface) features. High-symmetry directions used for the calculation of band dispersions are defined in panel d. **b,** Density of magnetization vector along high symmetry direction. Due to its complexity, we show only in-plane components perpendicular to *k*. The in-plane component parallel to wave vector is zero within the whole Brillouin zone. **c,** Schematic picture of the spin texture in main bulk bands extracted from panels a and b at *E*= -0.5 eV; the arrows denote the direction of the in-plane projection of magnetization vector for inner and outer bands. **a'-c',** Same as a-c for Ge-terminated surface, with polarization inwards. The Fermi level here has been shifted in order to align the bulk bands of Te- and Ge-terminated surfaces. **d,** Brillouin zone of hexagonal surface and bulk unit cells; the grey square marks the area displayed in panels c, c'. Solid vertical lines in **a, a'** indicate the *k* points used for the spin analysis reported in Figure 4.

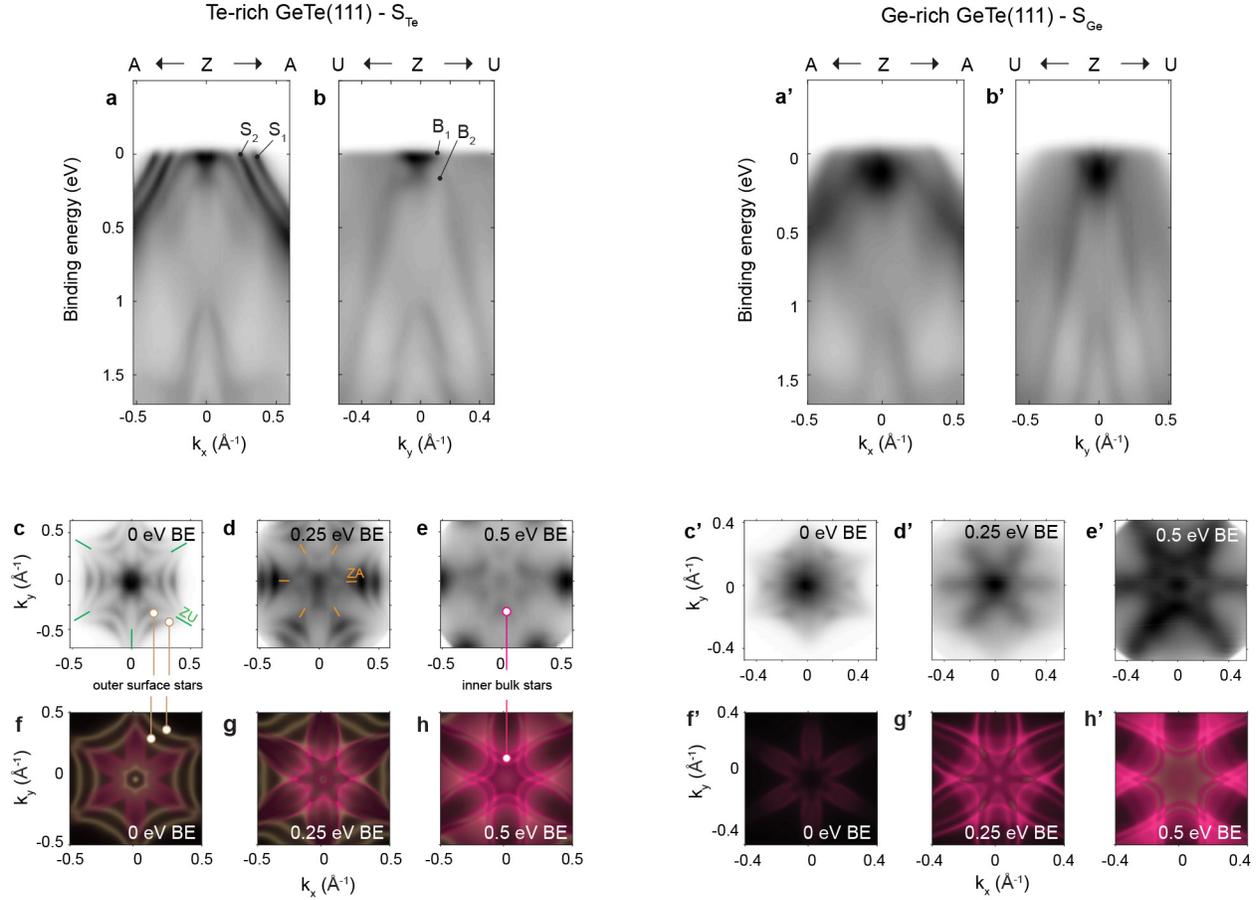

**Figure 3. ARPES from Te-rich and Ge-rich samples.** Panels **a-h** refer to the Te-rich sample $S_{Te}$ with outward polarization. **a, b** Experimental bands dispersion collected along the principal directions ZA ($k_x$) and ZU ($k_y$) in the BZ. **c, d, e** Constant energy maps at 0, 0.25 and 0.5 eV below the Fermi energy. **f, g, h** Corresponding calculated constant energy maps with yellow and pink indicating the surface and bulk character of states, respectively. Panels labelled by the apex (**a'-h'**) are the same as above but for the Ge-rich sample $S_{Ge}$ with inward polarization. The Fermi level in panels **f'-h'** is chosen consistently with Fig. 2.
16

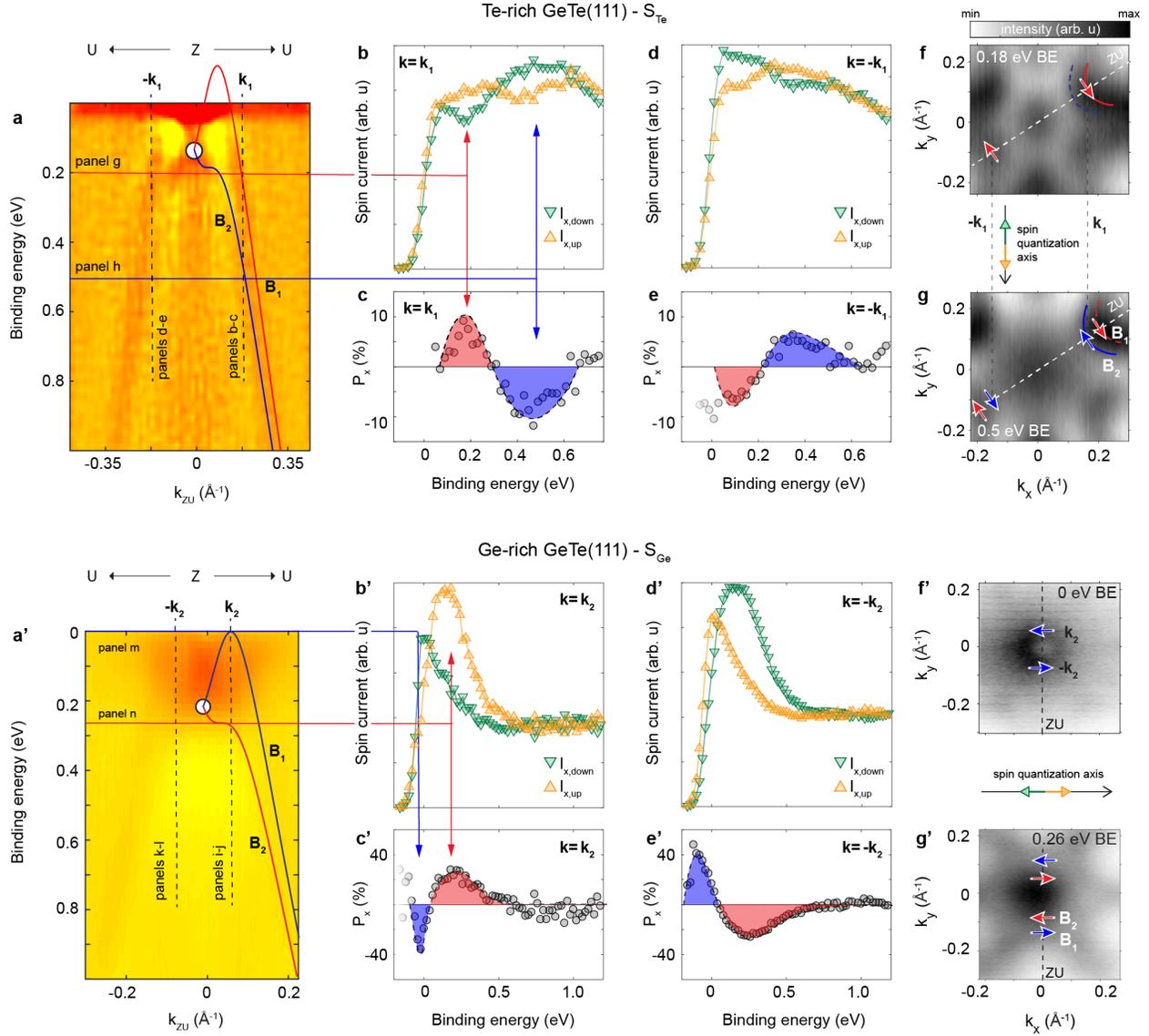

**Figure 4. Spin resolved ARPES from Te-rich and Ge-rich samples.** Panels **a-g** refer to the Te-rich sample $S_{Te}$ with outward polarization. **a,** Calculated bulk bands (solid line) along ZU ($k_y$) over the 2$^{nd}$ derivative of the measured band dispersion. **b, c,** Spin polarized spectra and spin asymmetry at fixed wave vector **k**$_1$ indicated in panel a. The two peaks correspond to the intersection of bulk Rashba bands B$_1$ and B$_2$ with the vertical dashed line at **k**$_1$ (panel a). **d, e,** Spin polarized spectra and spin asymmetry at opposite wave vector **-k**$_1$. **f, g,** Constant energy maps at 0.18 eV and 0.5 eV BE, corresponding to the energy of bulk bands B$_1$ and B$_2$ at **k**$_1$, in nice agreement with the peaks of opposite spin polarizations in c and e. Blue and red arrows indicate the sense of circulation of spins: clockwise in the outer band and counterclockwise in the inner one. Panels **a'-g'** refer to the case of Ge-rich sample $S_{Ge}$. **b', c', d', e'** Spin analysis for opposite wave vectors **k₂** and **–k₂** where the Rashba splitting is maximized. **f', g'** Constant energy maps at 0 eV (top of B1) and 0.26 eV BE, corresponding to the energy of bulk bands $B_1$ and $B_2$ at **k₂** (panel **a'**). The sense of circulation of spins is opposite to that found for $S_{Te}$: counterclockwise in the outer band and clockwise in the inner band (panel **g'**).